# Go local: The key to controlling the COVID-19 pandemic in the post lockdown era

Isabel Bennett[1*], Jobie Budd[1,2*], Erin M. Manning[1], Ed Manley[3], Mengdie Zhuang[4], Ingemar J. Cox[5,6], Michael Short[7], Anne M. Johnson[8], Deenan Pillay[9], Rachel A. McKendry[1,2]

1. London Centre for Nanotechnology, University College London; 2. Division of Medicine, University College London; 3. School of Geography, University of Leeds; 4. The Centre for Advanced Spatial Analysis, University College London; 5. Department of Computer Science, University College London; 6. University of Copenhagen; 7. Department for International Trade; 8. Institute of Global Health, University College London; 9. Division of Infection and Immunity, University College London;

* These authors contributed equally to this work.
Corresponding author: r.a.mckendry@ucl.ac.uk

> **Key recommendations:**
> - **A locality-based approach to lockdown easing is needed, enabling local public health and associated health and social care services to rapidly respond to emerging hotspots of infection.**
> - **National level data will hide an increasing heterogeneity of COVID-19 infections and mobility, and new ways of real-time data presentation to the public are required.**
> - **Data sources (including mobile) allow for faster visualisation than more traditional data sources, and are part of a wider trend towards near real-time analysis of outbreaks needed for timely, targeted local public health interventions.**
> - **Real time data visualisation may give early warnings of unusual levels of activity which warrant further investigation by local public health authorities.**

Abstract
The UK government announced its first wave of lockdown easing on 10 May 2020, two months after the non-pharmaceutical measures to reduce the spread of COVID-19 were first introduced on 23 March 2020. Analysis of reported case rate data from Public Health England and aggregated and anonymised crowd level mobility data shows variability across local authorities in the UK. A locality-based approach to lockdown easing is needed, enabling local public health and associated health and social care services to rapidly respond to emerging hotspots of infection. National level data will hide an increasing heterogeneity of COVID-19 infections and mobility, and new ways of real-time data presentation to the public are required. Data sources (including mobile) allow for faster visualisation than more traditional data sources, and are part of a wider trend towards near real-time analysis of outbreaks needed for timely, targeted local public health interventions. Real time data visualisation may give early warnings of unusual levels of activity which warrant further investigation by local public health authorities.

Main
The UK government announced its first wave of lockdown easing on 10 May 2020[1], two months after the non-pharmaceutical measures to reduce the spread of COVID-19 were first introduced on 23 March 2020. The easing represented a shift from the 'Stay at home' messaging of Phase one lockdown (delay, contain, research, mitigate) to 'Stay alert' messaging of Phase two (smarter controls). This was only applied to England in the first instance, with Wales, Scotland and Northern Ireland choosing to remain in Phase one with 'Stay at home' messaging and stronger mobility lockdown rules than England, such as a five mile local travel limit in Wales. The final Phase three (reliable treatment), will only be implemented once there is effective treatment and/or a vaccine. Since 10 May, there have been two subsequent lifting of restrictions on 1 and 15 June, with another announced for 4 July.

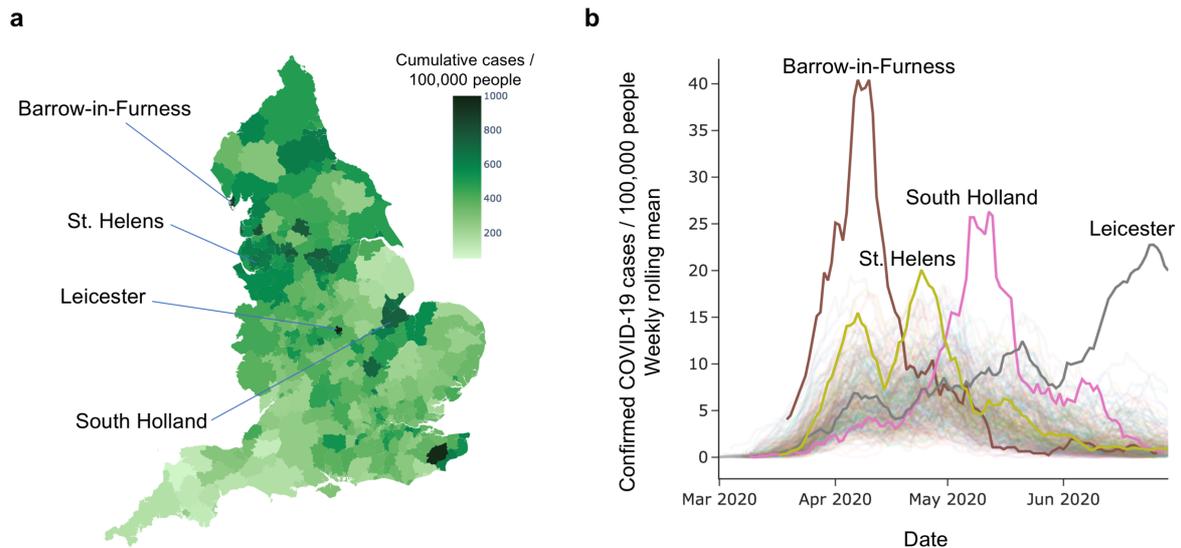

*Figure 1 a.* Cumulative reported COVID-19 case rate by local authority in England[2] from 30 January to 29 June 2020. Cases normalised by population in each local authority (per 100,000 people). *b.* Weekly rolling average of PHE daily reported COVID-19 case rate by local authority[2]. By way of illustration, infection peaks highlighted for four local authorities with the highest peak of weekly rolling average of daily cases per 100,000 population from 1 March to 29 June. Lockdown began in England on 23 March, and first easing was on 10 May.

A universal easing of lockdown in England was initiated despite significant COVID-19 infection burden across the country. Clusters of COVID-19 infection have been found across local authorities in England, as shown in Figure 1. Barrow-in-Furness (Cumbria) had a significant peak in reported infections at the beginning of April[3], with peaks appearing in several other local authorities in the following weeks, including St. Helens (Merseyside)[4], South Holland (Lincolnshire) and Leicester[5]. The use of real time data visualisation, for example Figure 1, may help local public health authorities identify unusual levels of activity earlier which warrant further investigation.

Anonymised and aggregated local authority level mobility data also show varying adherence to continued restrictions after the announcement of Phase two lockdown in England (Figure 2). Many areas across the UK returned to near-normal (pre-lockdown) levels of mobility on the announcement of the second wave of lockdown easing on 1 June (which included reopening of some schools and meetings of six people in open spaces allowed). However, some cities such as London, Manchester, Cambridge and Norwich, continue to show low levels of mobility, highlighting differences between mobility in rural and urban areas. The needs of these two populations is different, and intervention policies should reflect that where possible.

When comparing to Phase one lockdown levels of mobility (i.e. lowest levels), the data shows pockets of significant increases in activity since the easing (Figure 3). Whilst it is unsurprising that some of these hotspot areas would have increased mobility over the May Bank Holidays, such as the national parks (Peak District, Yorkshire Dales and Lake District) and coastal areas, it is worth highlighting the much higher levels of mobility in general across England compared to Scotland and Wales where stricter policies were maintained. This highlights the effect of policy on people's behaviour. Interestingly, mobility in Northern Ireland is comparable to England, despite having stricter policies in place[6].

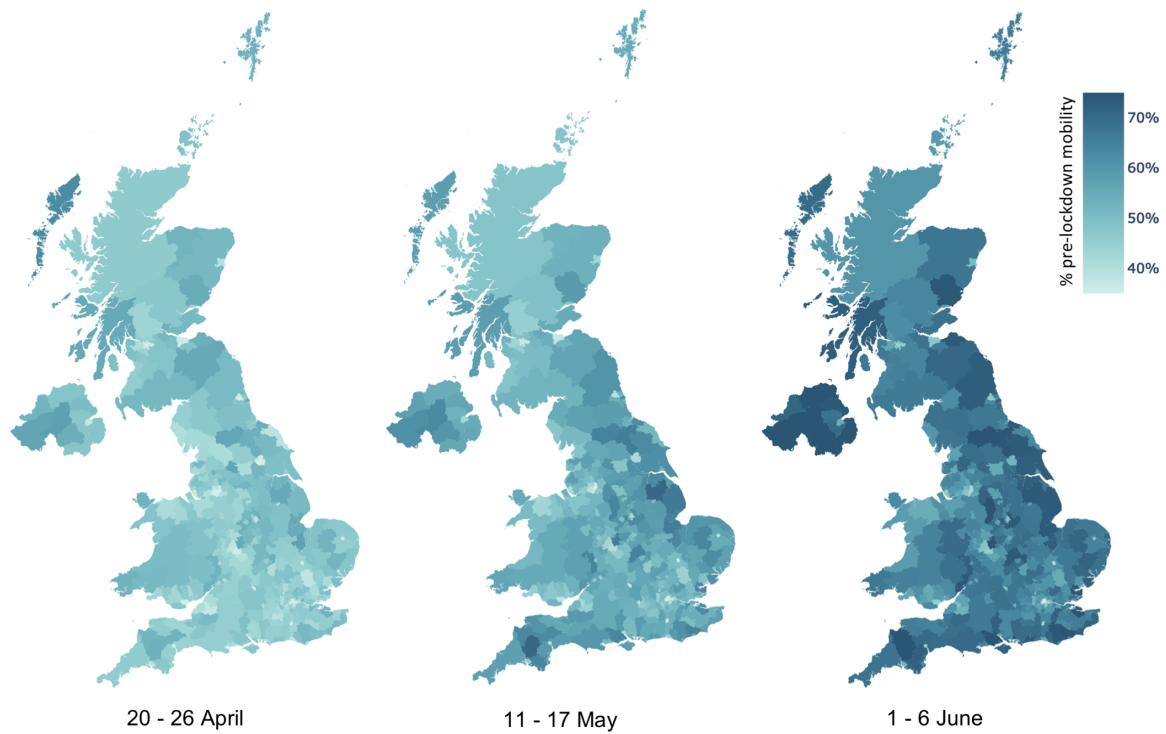

***Figure 2.*** *Effect of lockdown on mobility across the UK. Data shown during first week of lockdown Phase one (week of 20 April), week post first lockdown easing (week of 11 May), and week post second lockdown easing (week of 1 June) as percentage of pre-lockdown mobility (1 February – 7 March). Anonymised and aggregated crowd level mobility data provided by O2.*

Exactly how this increase in mobility to/within certain areas correlates with case clusters is a complex question, predominantly because of time delays with infection and onset of symptoms or testing[7,8]. Local case clusters could also be due to increased testing capacity in those areas, making interpretation of the data difficult. What this data does clearly show, however, is that lockdown mobility and case rates within England vary greatly across the local authorities. This may reflect availability of local facilities for a population, or the socioeconomic status, health, environmental and demographic risk factors of those living in the area – aspects which are lost when datasets are viewed in isolation from their local context. We believe a more nuanced and strategic approach to easing of lockdown measures, in consultation with local authorities, is needed and may be more effective than a universal approach.

The heterogeneity in mobility and confirmed cases suggests a need to decentralise national initiatives such as the proposed Joint Biosecurity Centre and Test and Trace system to locality level to facilitate rapid data linkage and resulting action. This will involve empowering local authorities to introduce temporary additional measures, such as controlling the opening of existing facilities, or creating additional facilities in locations where there are fewer shops or services, spreading the load of the local populations more evenly. This will allow for local differences in mobility needs and infection clusters to be quickly contained. This local approach has been introduced in France[9] and Germany[10], where differentiated measures were applied to local areas based on colour coded risk levels produced by key indicators. This data is presented as a publicly accessible dashboard with real-time data input.

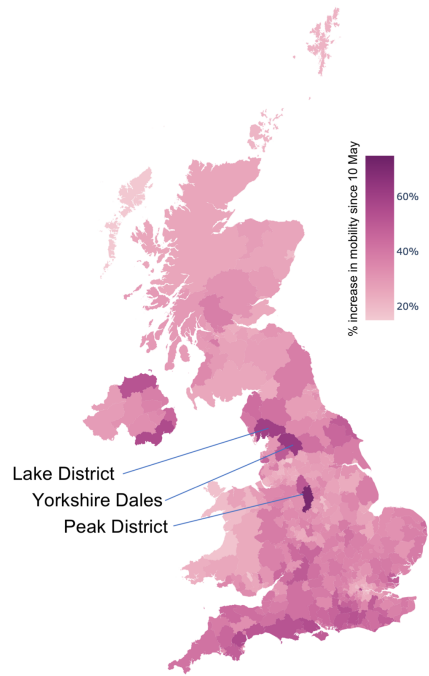

*Figure 3.* *Percentage increase in non-commute trips (see Data section for definition) between two weeks prior to easing on 10 May and first week of June by local authority. Data shows trips ending in local authority. Data for trips started and ending by local authority was found to be not significantly different. Anonymised and aggregated crowd level mobility data provided by O2.*

Finally, engaging with local authorities around their data will be important for communication and trust with the public. Co-development of policies to fit the local community needs will give people ownership and motivation to follow lockdown polices. Access to real-time data will be important to give these local policies credibility. This may involve engaging with industry partners who can provide more local level granularity of data. In addition, local authorities will vary in their access to resources and skills, with not all local authorities having established analysis teams which are well integrated with policymaking. Councils should work together on establishing plans and sharing insights, and making emergency resources available for key local authorities in the form of secondments or funds from central government.

This work supports the recommendation put forward in the recent Independent Sage report[11], specifically the recommendations for a local approach to containing COVID-19, empowering of local authorities, and use of data. This will allow for an agile and responsive system that can provide effective containment of the virus during this partial lockdown phase that England has entered.

## Data

*Anonymised and aggregated crowd level mobility data*

We received anonymised and aggregated, UK population representative crowd level mast data from O2 detailing the aggregate number of trips starting within each UK Lower Tier Local Authority (LTLA) for 1 February - 7 June 2020 inclusive[12]. LTLAs include County Districts (Non-Metropolitan Districts), Unitary Authorities, Metropolitan Districts and London Boroughs. Trips are created when anonymised devices move from one overlapping group of cells to another (where a cell is in an area of coverage provided by a single mobile network antenna) and remains with the group of cells for long enough to indicate the device is stationary.

Identifying commute/non-commute trips within mobility data: Where the start point of trips overlaps the general area of the anonymised user's regular home location in February 2020, and the end point overlaps the general area of the anonymised user's regular work location, or vice versa, trips will be flagged as commutes in the data. This definition is designed to identify regular commuting patterns, it may exclude shift workers, itinerant workers and other workers with unusual work patterns, or those who have changed their work location since February 2020.

*Cases data*

Cases data shows lab-confirmed cases from NHS and PHE laboratories ('pillar 1') and the mass testing program ('pillar 2'). Dates relate to the date of specimen collection. In the PHE dataset, duplicate tests for the same person are removed. The first positive specimen date is used as the specimen date for that person.

Confirmed positive cases are matched by PHE to Office for National Statistics (ONS) geographical area codes[13] using the home postcode of the person tested. Cases data is made openly available aggregated to LTLAs. Some cases cannot be matched to a geographical area by PHE because postcode information is missing or received late. Data for Hackney and City of London are combined and data for Cornwall and Isles of Scilly are combined. This is because City of London and Isles of Scilly have populations of less than 10,000 and publishing daily case numbers risks disclosing personal information about the outcome of COVID-19 tests.

The different LTLAs vary in population size. Areas with larger populations will tend to have more cases than those with smaller populations. To account for the different population sizes, rates are calculated by PHE. The count for each area is divided by the total population and multiplied by 100,000. Populations are the 2018 Mid-year Estimates from the Office for National Statistics[14].

Ethical Approval
Project ID 18255/001 (UCL Research Ethics Committee)

Acknowledgements
Anonymised and aggregated crowd level mobility data provided by O2.
This research was funded by i-sense: EPSRC IRC in Agile Early Warning Sensing Systems for Infectious Diseases and Antimicrobial Resistance EP/R00529X/1